\documentclass[12pt]{article}
\usepackage{graphicx}                                                                                                                             
\begin{document} \begin{center}

{\huge Desiccation and Relaxation Regimes in Fracture of Drying Clay Films}\\ \vskip 0.5cm
Dibyendu Mal$^1$, Suparna Sinha$^1$, Tapati Dutta$^2$, S. Mitra$^3$ and Sujata Tarafdar$^1$\\
$^1$Condensed Matter Physics Research Centre, Physics Department\\ Jadavpur University, Kolkata 700032, India\\
$^2$ Physics Department, St. Xavier's College, \\Kolkata 700016, India\\
$^3$ Department of Geological Sciences\\ Jadavpur University, Kolkata 700032\\India
\end{center}

\noindent {\bf Abstract} A study of crack patterns in laponite films of different thickness is presented. Two
cracking regimes are observed. The earlier with a higher rate of desiccation and shrinking has predominantly 
four-fold vertices
in the crack network, while the later regime has less volume shrinking and forms mainly three-fold vertices. The pattern
shows a self-similarity under coarse-graining. The graph of
area covered by the cracks versus minimum crack-width resolved, scales with the film thickness. Curves representing
crack area for different thickness can thus be made to collapse onto a single curve.

Crack patterns in drying clay films have a distinctive hierarchical structure, this has been discussed in several
earlier papers \cite{bohn,mal,shor}. Common features noted in the crack pattern in all these works are -  predominance of
four-sided peds and cracks meeting each other at right angles. However, even with these features in common, two
distinct classes of crack networks have been observed - (a) those with  three-fold vertices \cite{bohn} and (b) those with
predominantly four-fold vertices \cite{mal}. Class (a) is seen for example, in cracks on ceramic glazes, where the fractures
appear as hairline cracks. Though these form due to desiccation, the loss of fluid is not a significant fraction of the
total volume of the material. Class (b) on the other hand, is seen in dry mud flats, colloidal gels, or a suspension of
fine particles such as coffee powder, or laponite while losing solvent \cite{mal,shor,kaplan}. 
Here the cracks have a range of widths,
the largest may be comparable to the ped size, so a considerable volume shrinking is involved in the formation process.
The successively formed cracks, however, are narrower and narrower, so at some point of time the pattern 
becomes identical to the class (a) network, with the signature three-fold vertices. We call (b) the {\it desiccation regime}
and (a) the {\it relaxation regime}.
In this letter we present an experimental study of crack formation in drying laponite films and show the crossover
between the two distinct regimes. The difference between the two regimes has been identified by three  
characteristics - (i) different rates of weight loss during drying, (ii) predominance of either four-fold or three-fold 
vertices,
and (iii) different scaling rule with variation of film thickness for crack area as a function of crack width.

The experimental procedure is as follows - laponite (RD) is mixed with methanol and stirred for 1-2 hours to make a
suspension. The mixture is poured in a Petri-dish and allowed to dry. The crack pattern is photographed at intervals
as it forms. The Petri-dish is placed on a digital balance during drying, so the weight loss can be recorded continuously.

\begin{figure}[h]
\begin{center}
\includegraphics[width=9.0cm]{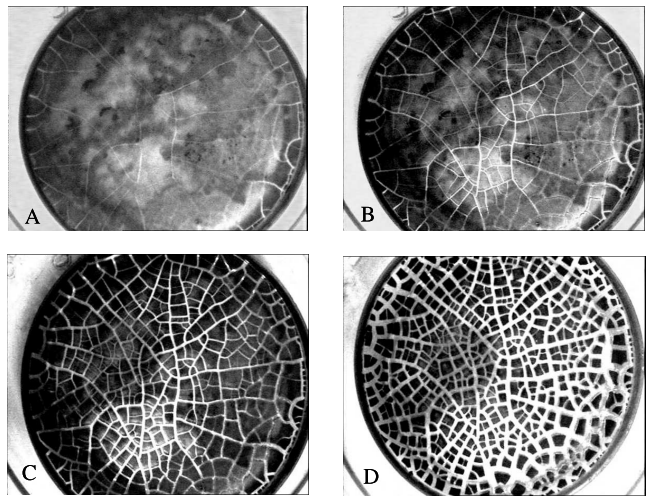}
\end{center}
\caption{A typical sequence of photographs showing the crack development with time.  } \label{photo}
\end{figure}

\begin{figure}[h]
\begin{center}
\includegraphics[width=9.0cm,angle=270]{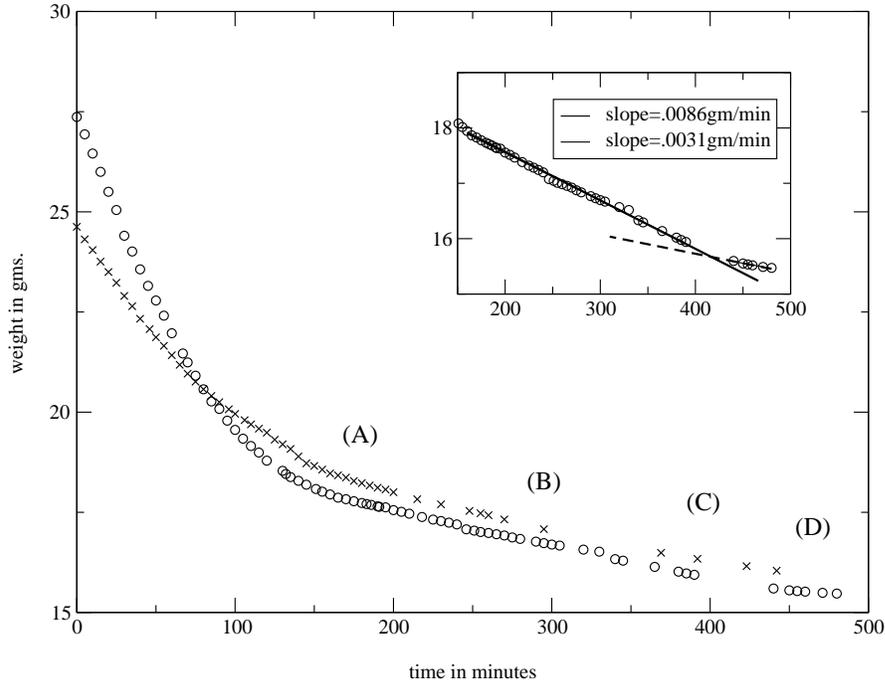}
\end{center}
\caption{Graph showing decrease in weight as the laponite film dries. The time when the photographs  
in figure(\ref{photo}) were taken are indicated as (A)-(D) on the graph. The initial steep slope is due to the excess fluid 
evaporating from the
saturated laponite-methanol mixture. The inset shows the change of slope from cracking regime (b), (A)-(C) in figure(\ref{photo}) to
(a), (C)-(D) in figure (\ref{photo}).} \label{wt}
\end{figure}

A typical sequence of
photographs is shown in figure (\ref{photo}). The weight as a function of time for two samples is shown in figure 
(\ref{wt}), positions (A)-(D) on the graph  indicate the time when the corresponding snapshots
in figure(\ref{photo}) were taken.
The initial rapid weight loss in figure \ref{wt} is due to excess methanol evaporating
from the top of the sample. The end of this regime is marked by appearence of white spots on the film surface.
We are henceforth concerned only with the region from (A)-(D) in figure (\ref{photo}).

\begin{figure}[h]
\begin{center}
\includegraphics[width=8.0cm]{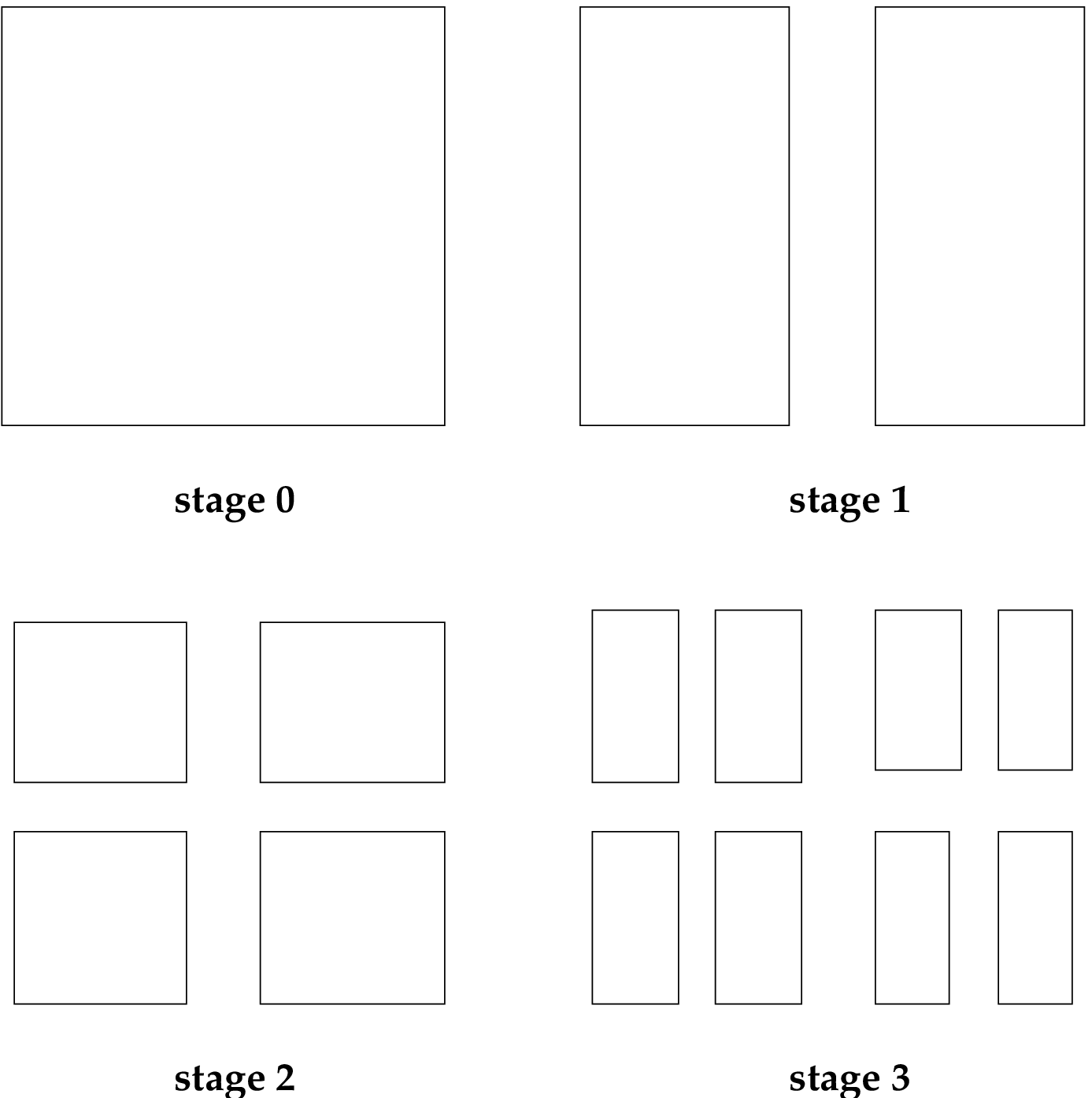}
\end{center}
\caption{The hierarchical sequence of crack formation and shrinking giving rise to the four-fold vertices.  } \label{pat1}
\end{figure}

The cracks first appear at the edge of the Petri-dish.                                                                       Figures (A)-(D), (\ref{photo}) show that the initial cracks are the longest and span 
the entire sample.The
segments formed shrink rapidly and subsequently these cracks are the widest. The next generation cracks form
when the segments again divide in a similar manner, these cracks are less wide. This hierarchical cracking process 
\cite{mal} is
illustrated schematically in figure (\ref{pat1}). The fractures appear near the centre of the segments where the stress
is maximal \cite{kom}. Four-way crossings or four-fold vertices arise in the crack network due to this sequence of
fracture. The time interval where this type of mechanism is operative is indicated in (A)-(C) in figure (\ref{wt}).

\begin{figure}[h]
\begin{center}
\includegraphics[width=6.0cm]{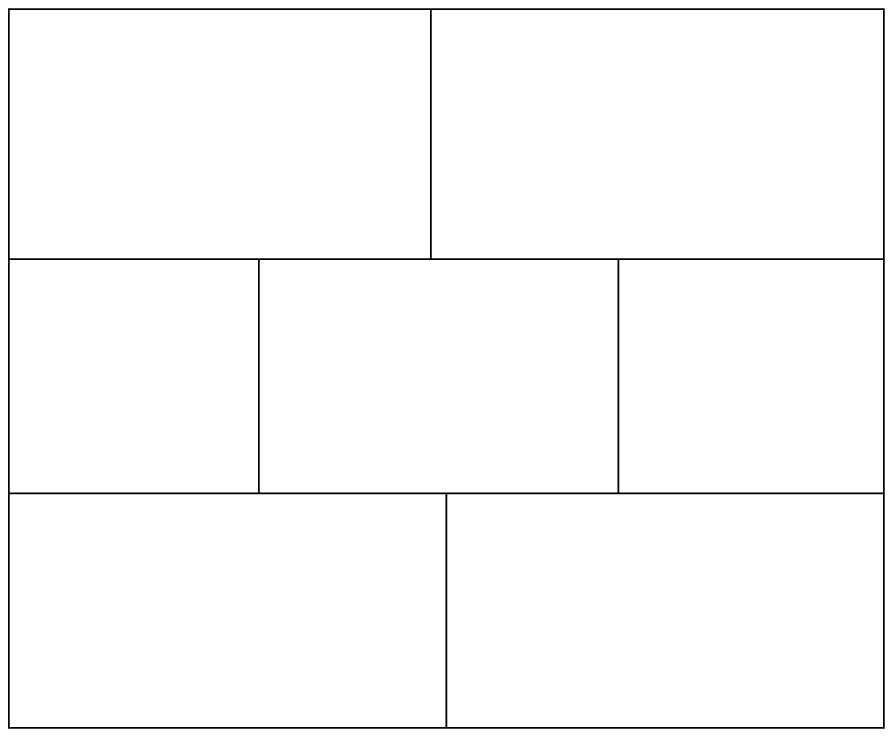}
\end{center}
\caption{ Schematic diagram for crack pattern in regime (a). } \label{pat2}
\end{figure}

At the point shown in the inset of figure (\ref{wt}), a change in slope of the drying rate can be observed. 
We suggest that this
is the point where a different mechanism takes over. Now the cracks are very narrow and desiccation is much less, the
cracks are formed as the clay particles relax and rearrange to relieve the residual stress. The narrow cracks meet at  
three-fold vertices, similar to
the pattern of laying bricks on a wall, the process is illustrated schematically in figure (\ref{pat2}). four-fold vertices
do not form because a second daughter crack does not meet the parent crack at a point where an earlier daughter crack has
already relieved the stress \cite{bohn}. This argument is not valid for the class (b) cracks, here the rapid shrinking makes
the  formed peds effectively independent of each other and the cracks appear at the middle. In regime (a) adjacent peds
are in contact and should be regarded as parts of the same system.

\begin{figure}[h]
\begin{center}
\includegraphics[width=7.0cm,angle=270]{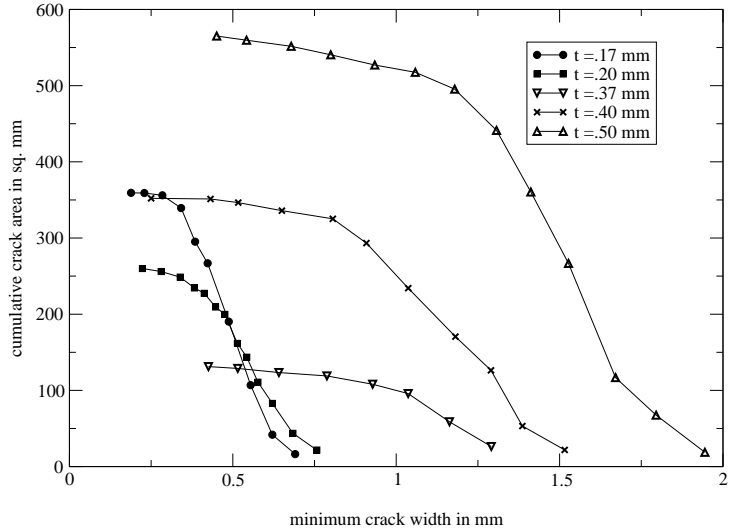}
\end{center}
\caption{ The crack patterns for films of different thickness are observed under different resolution. 
A plot of cumulative area vs. lowest observable crack width for various thicknesses is shown. } \label{area}
\end{figure}

In the second part of the analysis we concentrate on  
the self-similarity of the final pattern, manifest in the hierarchy of crack widths. When observed under slowly
increasing resolution i.e. decreasing the minimum observable crack width $w_{min}$ the curve representing the
cumulative area $A_{cum}$ 
covered by the cracks shows a typical highly reproducible shape. This curve again reveals the two different regimes, 
as discussed
earlier in \cite{mal}. We show in figure \ref{area} the curves for a range of film thicknesses $t$. 
The high resolution (low $w_{min}$) side of these curves all show the class (a) regime, where crack width is negligible
and  $A_{cum}$ changes very slowly. The low resolution (high $w_{min}$) side can be explained on the basis of 
the formation process in 
figure \ref{pat1} with a successive decrease in width of higher stage cracks by a factor of about $0.93$ \cite{mal}.
So in systems involving desiccation of a large amount of solvent, both the
processes described by figures (\ref{pat1},\ref{pat2}) are found, and the resulting network has a combination of 
three and four fold vertices.
\newpage

\begin{figure}[h]
\begin{center}
\includegraphics[width=7.0cm,angle=270]{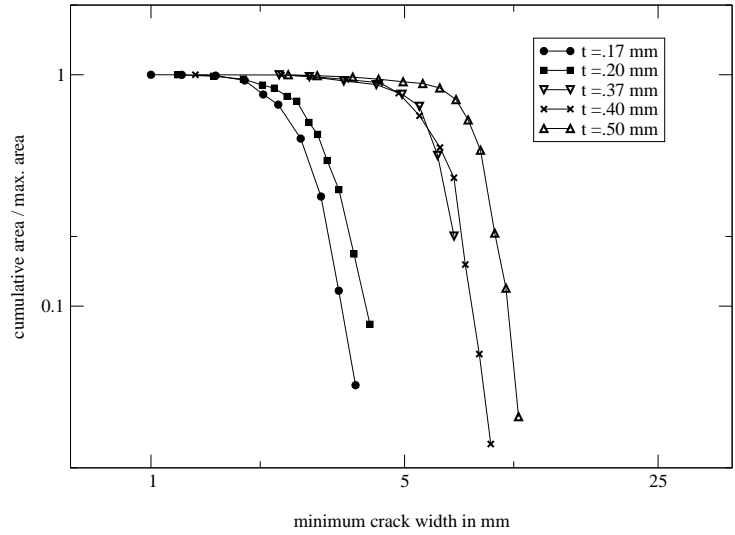}
\end{center}
\caption{ Fractional crack area vs. minimum crack width for various thicknesses  } \label{sca1}
\end{figure}

\begin{figure}[h]
\begin{center}
\includegraphics[width=7.0cm,angle=270]{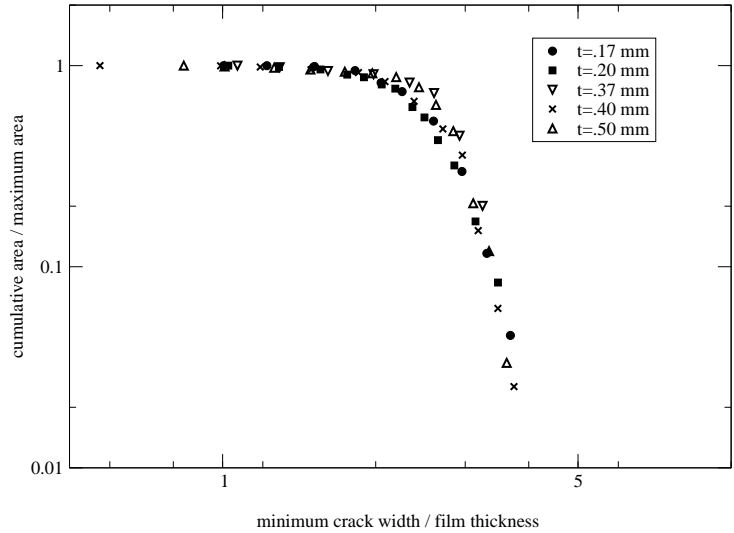}
\end{center}
\caption{ Fractional crack area vs. minimum crack width scaled by the film thickness. All the curves for
different $t$ collapse onto a single curve. } \label{sca2}
\end{figure}

A very  interesting feature of the two cracking regimes is their different scaling behaviour with variation in
thickness $t$ of the film. If we plot the data shown in figure (\ref{area}), as the fractional
crack area (dividing $A_{cum}$ by the maximum crack area), we find that the narrow crack regimes for all the
curves merge
showing that here the functional dependence of $A_{cum}$ on $W_{min}$ is independent of the film thickness. This is shown in figure (\ref{sca1}). Here $A_{cum}$ is very weakly dependent on $W_{min}$.
It may be noted that the total sample area for all these patterns are not equal. In some samples small regions of very
different thickness due to inhomogeity of the suspension had to be omitted during the analysis.

If we now scale the crack width (the x-axis) by the film thickness $t$, we find that all the curves collapse quite well onto
a single curve figure \ref{sca2}. This shows that for the wider crack regime 
\begin{equation} A_{cum} = F(W_{min}/t) \end{equation} $F$ being some function.

For the widest cracks, this regime follows a power-law with an exponent close to 8. 
It was shown in \cite{mal} that this corresponds
to  cracks of successive stages in the pattern of figure (\ref{pat2}) being reduced in width by a factor of 0.93.

The scaling of the cumulative area curves with $t$ can be explained qualitatively as follows - we assume following  
\cite{kaplan}, that the cracking
is due to the friction between the surface of the Petri-dish and the laponite-methanol mixture, which resists the
shrinkage and this
force is very weakly dependent on the thickness of the drying layer. The energy gained by shrinking of the mixture
is more for the thicker samples in proportion to their thickness, so they can shrink more creating wider cracks
before the gain in energy due to shrinking is balanced by the pinning frictional force. Our results are consistent
with the observation that the linear size of the crack patterns scales with film thickness \cite{kaplan}.
 
To quanitify our analysis further, we count the three and four-fold  vertices in several patterns, the numbers compare as
shown in table 1. 
\begin{table}
\begin{center}
\begin{tabular}{|c|c|c|} \hline
Sample & no. of 4-fold vertices & no.of 3-fold vertices \\ \hline
(a)  & 63 & 102 \\
(b) & 78  & 180 \\
(c) & 86  & 200  \\ \hline \end{tabular}
\caption{ The number of 3-fold and 4-fold vertices counted for three different patterns} \end{center}
\end{table}

A very rough estimate of the number of vertices can be made on the basis of the scheme shown in
figure (\ref{pat1}). During earlier part of the crack formation regime (b), each successive stage increases the number of
segments by a factor of 2, and two successive stages are required to create four new four-point vertices. 
We can estimate the number of stages $n$ in the (b) regime from the ratio of maximum crack width for each $t$ -  
$W_{max}$ to $w_{min}$ at the cross-over from
regime (b) to (a). This ratio is seen from figure (\ref{area}) to be close to 0.6 for all the samples. Since each stage
reduces the crack width by 0.93, we have
\begin{equation}
w_{max}/w_{min}(crossover) = 0.6 = 0.93^n
\end{equation}
$n$ from such an estimate is found to be 7. According to figure (\ref{pat1}) starting with 1 four-fold vertex at stage 2
this should produce $4^3 = 64$  four-fold vertices. This is of course, an approximate `order of magnitude' estimate, the
irregularities in the substrate and inhomogeneity of the laponite suspension make the process highly stochastic.
Moreover boundary effects due to the circular geometry of the periphery of the Petri-dish creates further disorder 
in the initial
stages (see figure (A)-(B) (\ref{photo}). 

The regime (a) which
follows creates the three-fold vertices. Here the number of peds created in each stage is not well defined, but assuming 
three or four daughter peds produced from each ped after completion of regime (b), 
an increase by a factor of 2  in the number of new three-fold
vertices, as observed in Table 1 is quite reasonable.

 In spite of the uncertainties involved, the
reproducibility of the overall features and the existence of the two different cracking processes seems quite unmistakable. 
In figure (\ref{area}) the crossover from regime (b) to (a) occurs at around $$W_{min} (cross over) = 0.6 W_{max}$$ for
all $t$. The cracks with width $W_{min}(cross over)$ contribute most to the total area covered by the cracks.

Earlier work on crack-patterns deals mainly with ped size, shape and their relation with film thickness \cite{bohn,kaplan}.
Studies involving width of cracks and their distribution are lacking.
Besides the studies on hierarchical crack patterns a whole body of literature deals with directional drying and patterns
it creates e.g.\cite{allain,kom}. The geometry of such systems is somewhat different from ours. The fractal dimensions
of branching cracks have also been studied \cite{wang,bou}.

We have done this study on laponite as it is a synthetic clay with monodisperse nanosized particles. It has  properties
similar to natural clays but is free from heterogeneity and impurities in real clay and the structure is well known. 
Methanol is used to make the dispersion as it
evaporates quickly, but being non-polar, it does not form a gel with laponite. 
We plan to study also the patterns of laponite gels
using water as the solvent.

We hope this analysis may provide a basis for
more quantitative studies in future and lead to a better understanding of the important phenomenon of crack formation.                                                                                                                              
Authors are thankful to Connell Brothers (India) and Rockwood Additives (UK) for providing the Laponite(RD) samples free
of cost. DM thanks UGC for providing a research
fellowship under project No. F5-9/2003(SR). We are grateful to Surajit Sengupta for stimulating 
discussion on pattern formation.
  
\end{document}